\documentclass[lettersize,journal]{IEEEtran}
\IEEEoverridecommandlockouts
\usepackage{amsmath}
\usepackage{cite}
\usepackage{amsmath,amssymb,amsfonts}
\usepackage[ruled,linesnumbered,lined]{algorithm2e}
\usepackage{graphicx}
\usepackage{textcomp}
\usepackage{xcolor}
\usepackage{bm}
\usepackage{booktabs}
\usepackage{amsthm,amssymb,amsfonts}

\usepackage{float}
\usepackage{subfig}

\newtheorem{theorem}{Theorem}

\newtheorem{lemma}{Lemma}

\SetKwInput{Input}{input}
\SetKwInput{Output}{output}
\newtheorem{definition}{Definition}
\usepackage{array}
\newcolumntype{L}[1]{>{\raggedright\let\newline\\\arraybackslash\hspace{0pt}}m{#1}}
\newcolumntype{C}[1]{>{\centering\let\newline\\\arraybackslash\hspace{0pt}}m{#1}}
\newcolumntype{R}[1]{>{\raggedleft\let\newline\\\arraybackslash\hspace{0pt}}m{#1}}

\newcommand{\T}{\mathbf \Theta}

\def\BibTeX{{\rm B\kern-.05em{\sc i\kern-.025em b}\kern-.08em
		T\kern-.1667em\lower.7ex\hbox{E}\kern-.125emX}}

\begin{document}
	\title{Joint Active and Passive Beamforming Optimization for Beyond Diagonal RIS-aided Multi-User Communications}
	\author{Xiaohua Zhou, \textit{Graduate Student Member, IEEE}, Tianyu Fang, \textit{Graduate Student Member, IEEE}, Yijie Mao, \textit{Member, IEEE}
		\thanks{This work has been supported in part by the National Nature Science Foundation of China under Grant 62201347; and in part by Shanghai Sailing Program under Grant 22YF1428400.
			\par X. Zhou and Y. Mao are with the School of Information Science and Technology, ShanghaiTech University, Shanghai 201210, China (e-mail:
			{zhouxh3, maoyj}@shanghaitech.edu.cn). (\textit{Corresponding Author: Yijie Mao})
			\par Tianyu Fang is with Centre for Wireless Communications, University of Oulu, Finland (e-mail: tianyu.fang@oulu.fi).
		}
		\vspace{-0.18cm} 
		
	}

	\maketitle

	\thispagestyle{empty}
	\pagestyle{empty}
	\begin{abstract}
		Benefiting from its capability to generalize existing reconfigurable intelligent surface (RIS) architectures and provide additional design flexibility via interactions between RIS elements, beyond-diagonal RIS (BD-RIS) has attracted considerable research interests recently. However, due to the symmetric and unitary passive beamforming constraint imposed on BD-RIS, existing joint active and passive beamforming optimization algorithms for BD-RIS either exhibit high computational complexity to achieve near optimal solutions or rely on heuristic algorithms with substantial performance loss. In this paper, we address this issue by proposing an efficient optimization framework for BD-RIS assisted multi-user multi-antenna communication networks. Specifically, we solve the weighted sum rate maximization problem by introducing a novel beamforming optimization algorithm that alternately optimizes active and passive beamforming matrices using iterative closed-form solutions. Numerical results demonstrate that our algorithm significantly reduces computational complexity while ensuring a sub-optimal solution.
	\end{abstract}
	
	\section{Introduction}
	
	Recently, beyond-diagonal reconfigurable intelligent surface
	(BD-RIS) has been proposed in \cite{shen2021modeling} as a pioneering RIS architecture that provides notable performance gains in wireless communications. Pursuant to the microwave theory, BD-RIS introduces a beyond-diagonal scattering matrix (also known as passive beamforming matrix), allowing for interconnections among its elements. According to different interconnection methods among the elements, BD-RIS can be categorized into: single, group, and fully connected architectures \cite{hongyu2023}. 

	\par Realizing the performance benefits provided by BD-RIS requires careful passive beamforming design. In point-to-point communications, the optimal and efficient passive beamforming design for BD-RIS has been explored to maximize the channel gain \cite{nerini2023closed, Santamaria2023,fang2022fully,demir2024wideband}. However, when considering multi-user communications, maximizing the channel gain only is insufficient for optimizing system performance. This is because multi-user interference brings extra beamforming design challenges. A typical approach is to consider joint optimization of active beamforming at the transmitter and passive beamforming at the BD-RIS.  For the classical weighted sum-rate (WSR) maximization problem, various joint beamforming optimization algorithms have been proposed \cite{shen2021modeling,fang2022fully,hongyu2023,hongyu2024,hongyu2023jasc,yuyuan2023}, which typically involve an alternative optimization (AO) algorithm to iteratively optimize active and passive beamforming. The primary difference among these algorithms lies in the approach  to solve the passive beamforming subproblem. 
 
 \par The earliest approach to optimize the BD-RIS passive beamforming matrix is the quasi-Newton method \cite{shen2021modeling,fang2022fully}. However, as the number of  BD-RIS elements increases, the corresponding computational complexity rises sharply due to vectorization for the passive beamforming, resulting in an unacceptable computational burden. To address this issue, a manifold algorithm is proposed \cite{hongyu2023,hongyu2024,hongyu2023jasc,yuyuan2023}. This method relaxes the symmetric unitary matrix constraint to a unitary constraint. Despite attempts to project the solution to preserve the symmetric unitary matrix constraint, there is a non-negligible loss in performance. Furthermore, a passive beamforming optimization algorithm based on penalty dual decomposition (PDD) was introduced in \cite{yuyuan2023} for BD-RIS. Within the PDD framework, although a closed-form solution is derived, the computational complexity remains high. This is primarily due to the vectorization of the passive beamforming matrix into a high-dimensional vector and the inclusion of an inverse operation on a high-dimensional matrix within the derived closed-form expression. To reduce the computational complexity, an efficient passive beamforming design algorithm has been proposed in our previous work \cite{fang2023low}, which provides a heuristic two-stage solution to the WSR maximization problem. Although such algorithm performs well in small-scale networks, the performance loss becomes noticeable when dealing with large number of transmit antennas and users.

	\par In this work, we aim to address all the aforementioned limitations of existing joint active and passive beamforming design algorithms for a BD-RIS-aided multi-user downlink transmission network. We propose an efficient optimization algorithm that guarantees the  near optimal solution for the original WSR maximization problem while significantly reducing the computational time. The beauty of the proposed algorithm lies in a unified optimization framework, which employs iterative closed-form solutions to address both active and passive beamforming design subproblems. Importantly, this method eliminates the need for optimization toolboxes, resulting in a notable reduction in computational complexity. Simulation results show that compared with existing baselines, the proposed optimization framework substantially reduces computation time while maintaining near optimal WSR performance in large-scale networks.


	\section{System  Model and Problem Formulation}\label{Sec:system model}
	
	
	Consider a BD-RIS assisted multi-user downlink communication network comprising a base station (BS) with $ L $ antennas, a BD-RIS with $ N $ passive reflecting elements, and a set of $ K $ users, each with a single receive antenna. The user index set is $ \mathcal{K} = \{1, \cdots, K\} $. The BS serves all $ K $ users simultaneously with the assistance of the BD-RIS. Let $ s_k\in\mathbb{C} $ denote the data symbol intended for user $ k $ and $ \mathbf w_k\in\mathbb C^{L\times 1} $ represent the corresponding active beamforming vector at the BS. The data symbol vector is $  \mathbf s \triangleq [ s_1,\cdots, s_K]^T \sim\mathcal{CN}(\mathbf 0,\mathbf I_K) $. The resulting transmit signal at the BS is $ \mathbf{x}=\sum_{k=1}^{K}\mathbf w_k  s_k. $

	The signal is transmitted through the BD-RIS aided channel, assuming that the direct links from the BS to the users are blocked. Let $ \bm \Theta \in\mathbb C^{N\times N}$ denote the passive beamforming matrix of the BS-BIS. Also, we denote the channel between the BD-RIS and user $ k $ as $ \mathbf h_k\in\mathbb C^{N\times 1} $, and the channel between the BS and BD-RIS as $ \mathbf E\in\mathbb C^{N\times L} $. The effective channel between the BS to user $k$ is therefore denoted as $\mathbf{f}_k^H= \mathbf h_k^H\bm\Theta\mathbf E$. The resulting receive signal at user $ k $ is:
	\begin{equation}\label{eq:received_signal}
		y_k=\mathbf f_k^H\sum\nolimits_{i=1}^{K} \mathbf w_i s_i+ n_k,
	\end{equation}
	where $ n_k\sim\mathcal{CN}( 0,\sigma_k^2) $ represents the additive white Gaussian noise (AWGN). 
	
	\par By further defining $ \mathbf y\triangleq[ y_1,\cdots, y_K]^T\in\mathbb C^{K\times 1}$, $ \mathbf H\triangleq [\mathbf h_1,\cdots,\mathbf h_K]\in\mathbb C^{N\times K} $, $ \mathbf W\triangleq [\mathbf w_1,\cdots,\mathbf w_K]\in\mathbb C^{L\times K} $ and $ \mathbf n\triangleq[ n_1,\cdots, n_K]^T\in\mathbb C^{K\times 1} $, the overall signal received at all users can be written in a compact form as:
	\begin{equation}
		\mathbf y= \mathbf H^H\mathbf \Theta\mathbf E\mathbf W\mathbf s+\mathbf n.
	\end{equation} 
	
	\par 
	The achievable rate at user $k$  is given as:
	\begin{equation}\label{eq:rate}
		\begin{aligned}
			R_k= \log \left (1+ \frac{|\mathbf f_k^H\mathbf w_k|^2}{\sum_{j=1,j\neq k}^K |\mathbf {f}_k^H \mathbf w_j|^2+\sigma_k^2  }  \right).
		\end{aligned}
	\end{equation}

	In this work, we focus on the fundamental WSR maximization problem subject to the total transmit power constraint at the BS, i.e., $ \|\mathbf W\|^2_F \leq  P_t $. It has been shown in \cite{fang2024ratesplittingmultipleaccess,rethingkingWMMSE} that such transmit power constraint is equivalent to $ \|\mathbf W\|^2_F =  P_t $ for the WSR problem. This leads to the following feasible set of the transmit beamforming matrix $\mathbf{W}$:
	\begin{equation}\label{eq:power-constraint}
		\mathcal S \triangleq\{\mathbf W\in\mathbb{C}^{L\times K}\big| \|\mathbf W\|^2_F =  P_t  \}.
	\end{equation}
	In this study, we consider a fully connected BD-RIS \cite{shen2021modeling}. It consists of a reconfigurable impedance network, where each port is interconnected with all other ports through reconfigurable impedances. Consequently, the passive beamforming matrix $ \mathbf \Theta $  belongs to the following manifold:
	\begin{equation}\label{eq:theta-constraint}
		\mathcal M \triangleq \big\{\bm\Theta \in\mathbb C^{N\times N}\big| \bm\Theta=\bm\Theta^T, \bm\Theta\bm\Theta^H=\mathbf I_N\big\}.
	\end{equation}


	
 Let $\delta_k$ denote the weight of user $k$, the formulated WSR maximization problem is given as:
	\begin{equation}\label{P1}
		\max_{\bm\Theta\in\mathcal M, \mathbf W\in\mathcal S}\,\, \sum_{k=1}^K \delta_k R_k.
	\end{equation}
	\par Solving problem \eqref{P1} poses significant challenges since the active and passive beamforming matrices are coupled in the fractional signal-to-interference-plus-noise ratios (SINR) expressions, and the symmetric and unitary constraint is highly non-convex. Conventional  optimization approaches, such as the quasi-Newton method \cite{shen2021modeling,fang2022fully}, the manifold method \cite{hongyu2023,hongyu2024,hongyu2023jasc}, the PDD method \cite{yuyuan2023}, and the two-stage algorithm \cite{fang2023low} have been proposed to solve problem \eqref{P1}. However,  they are all subject to the limitations outlined in the introduction section. In the next section, we propose an efficient optimization framework to resolve these challenges.

	\section{Proposed Optimization Framework}
	In this section, we propose an efficient AO method to solve problem \eqref{P1} by iteratively optimizing the active and passive beamforming matrices until convergence. Unlike conventional methods, such as quasi-Newton and manifold, which rely on optimization toolboxes to solve the subproblems of the active and passive beamforming matrices iteratively, the primary novelty of the proposed method is its use of iterative closed-form solutions for both subproblems. This substantially reduces computational complexity. The following subsections detail the proposed optimization algorithm.
	\subsection{Fractional Programming Reformulation}\label{FP}
	
	\par In order to transform problem \eqref{P1} into a more tractable form, we first apply the fractional programming (FP) framework \cite{shen2018fractional}, leading to the following lemma.
	\begin{lemma}
		By introducing auxiliary variables $ \bm \alpha =\{\alpha_{1}, \cdots, \alpha_K \}, \bm \beta=\{\beta_1, \cdots, \beta_K\} $, problem \eqref{P1} is equivalently reformulated as:
		\begin{equation}\label{P4}
			\max_{\bm\alpha,\bm\beta,\mathbf W\in\mathcal S,\bm\Theta\in\mathcal M }\,\, \sum\nolimits_{k=1}^K\delta_k h_k(\alpha_k,\beta_k,\mathbf W,\bm\Theta),	
		\end{equation}
		where
		\begin{equation}\label{eq:h-k}
			\begin{aligned}
				h_k(&\alpha_k,\beta_k,\mathbf W,\bm\Theta)=2\sqrt{1+\alpha_k} \Re\{\beta_k^*\mathbf f_k^H\mathbf w_k \}-\alpha_k\Bigg.\\
				& \Bigg.-|\beta_k|^2\left(  \sum\nolimits_{j=1}^K|\mathbf f_k^H\mathbf w_j|^2+\sigma_k^2 \right)+\log(1+\alpha_k).
			\end{aligned}
		\end{equation}
	\end{lemma}
	\textit{Proof:}
	The equivalence has been readily established by applying the Lagrangian dual and quadratic transform proposed in \cite{shen2018fractional}. Specifically, by introducing an auxiliary variable $ \alpha_k $, the logarithmic rate expression $R_k$ is reformulated as:
	\begin{equation}\label{ldt_alpha}
		R_k=\max_{\alpha_k}\,\, \log(1+\alpha_k)-\alpha_k+\frac{(1+\alpha_k)|\mathbf f^H_k\mathbf w_k|^2 }{\sum_{j=1}^K|\mathbf f_k^H\mathbf w_j|^2+\sigma_k^2}. 
	\end{equation}
	By further introducing another auxiliary variable $ \beta_k $ to decouple the fractional term in \eqref{ldt_alpha}, $R_k$ can be further recast as:
	\begin{equation}\label{ldt_beta}
		\begin{aligned}
			R_k=\max_{\alpha_k,\beta_k}\,\,	h_k(\alpha_k,\beta_k,\mathbf W,\bm\Theta). 
		\end{aligned}
	\end{equation}
	By plugging the reformulated achievable rate into problem \eqref{P1}, we obtain problem \eqref{P4}. 
	\par With fixed $ \mathbf W ,\bm\Theta$, the subproblems with respect to $ \alpha_k $ or $ \beta_k $ are unconstrained convex problems. By setting their partial derivative as zero, the optimal solutions of $ \alpha_k $ and $ \beta_k $ are:
	\begin{subequations}\label{eq:optalphabeta}
		\begin{align}
			\label{eq:optalpha}	\alpha_k^\star&=\frac{|\mathbf f_k^H\mathbf w_k|^2}{\sum_{j=1,j\neq k}^K |\mathbf f_k^H\mathbf w_j|^2+\sigma_k^2},\forall k\in\mathcal K,\\
			\label{eq:optbeta}	\beta_k^\star&=\frac{\sqrt{1+\alpha_k}\mathbf f_k^H\mathbf w_k}{\sum_{j=1}^K |\mathbf f_k^H\mathbf w_j|^2+\sigma_k^2},\forall k\in\mathcal K.
		\end{align}
	\end{subequations} 
	By substituting \eqref{eq:optalphabeta} into \eqref{eq:h-k}, we obtain \eqref{eq:rate}. Hence, the transformation from problem \eqref{P1} to \eqref{P4} is equivalent.
	$\hfill\blacksquare$

	Problem \eqref{P4} is a block-convex problem.
    To derive a near optimal solution, we iteratively update $\{\bm \alpha, \bm \beta\}$, $\mathbf{W}$, and $\mathbf{\Theta}$. Note that with fixed $\mathbf{W}$ and $\mathbf{\Theta}$, the optimal solutions of $\{\bm \alpha, \bm \beta\}$ are obtained by \eqref{eq:optalphabeta}. In the following, we illustrate a novel unified optimization framework with iterative closed-form solutions for the subproblems of $\mathbf{W}$ and $\mathbf{\Theta}$.	
	
	\subsection{Passive Beamforming Optimization}\label{FP_Theta}
	
	With given $\bm\alpha,\bm\beta, \mathbf W $, the subproblem of \eqref{P4} with respect to $ \bm\Theta $ becomes:
	\begin{equation}\label{eq:sub-theta}
		\max_{\bm\Theta\in\mathcal M }\,\, \sum_{k=1}^K2 \sqrt{1+\alpha_k} \delta_k \Re\{\beta_k^*\mathbf f_k^H\mathbf w_k \}
		-\delta_k|\beta_k|^2 \sum_{j=1}^K|\mathbf f_k^H\mathbf w_j|^2,	
	\end{equation}
	where $\mathbf{f}_k^H= \mathbf h_k^H\bm\Theta\mathbf E$.
	Through the definition of the following matrix variables:	
	\begin{equation*}
		\begin{aligned}
			&	\mathbf M\triangleq\mathbf E\mathbf W\mathbf\Sigma_1\mathbf H^H, \mathbf Y\triangleq \mathbf H\mathbf \Sigma_2\mathbf H^H,\\
			& \mathbf X\triangleq\mathbf E\mathbf W\mathbf W^H\mathbf E^H, \mathbf \Sigma_2\triangleq\mathrm{diag}\{\delta_1|\beta_1|^2,\cdots,\delta_K|\beta_K|^2 \}, \\
			& \mathbf \Sigma_1\triangleq\mathrm{diag}\{\delta_1\sqrt{1+\alpha_1}\beta_1,\cdots,\delta_K\sqrt{1+\alpha_K}\beta_K \},
		\end{aligned}
	\end{equation*}
	we equivalently transform \eqref{eq:sub-theta} into:
	\begin{equation}\label{P4Theta}
		\max_{\bm\Theta\in\mathcal M}\,\, 2\Re\{\mathrm{tr}\left( \bm\Theta \mathbf M \right) \}-\mathrm{tr}\left(\mathbf \Theta\mathbf X\mathbf \Theta^H\mathbf Y \right).  
	\end{equation}
	
	Although the objective function of \eqref{P4Theta} is concave, it remains challenging to solve due to the highly non-convex constraint of $\bm \Theta$. To address \eqref{P4Theta}, we first introduce a constant term $ \operatorname{tr}(\rho_1 \mathbf X) $ to transform the objective function into a convex form, where $ \rho_1 $ is an arbitrary constant such that $ \rho_1\mathbf I_N-\mathbf Y $ is a positive semi-definite matrix \cite{golub2013matrix}. With $ \mathbf \Theta \mathbf \Theta^H = \mathbf I_N $ in \eqref{eq:theta-constraint},
	 \eqref{P4Theta} is transformed into a more tractable form as:	 
	\begin{equation}\label{P4Thetas}
		\max_{\bm\Theta\in\mathcal M}\,\, 2\Re\{\mathrm{tr}\left( \bm\Theta \mathbf M \right) \}+\mathrm{tr}\left(\mathbf \Theta\mathbf X\mathbf \Theta^H(\rho_1\mathbf I_N-\mathbf Y) \right).  
	\end{equation}

	To solve problem \eqref{P4Thetas}, we introduce the following lemma.
	\begin{lemma}\label{SCA_matrix}
		Let $ \mathbf A,\mathbf B\in\mathbb{C}^{N\times N} $ be two positive semi-definite Hermitian matrices, then for any matrices $ \bm\Theta'\in\mathbb{C}^{N\times N},\bm\Phi'\in\mathbb{C}^{N\times N} $, we have
		\begin{equation}\label{eq:linear-theta}
			\mathrm{tr}(\T'\mathbf A\T'^H\mathbf B)\geq 2\Re\{\mathrm{tr}(\bm\Phi'\mathbf A\mathbf \Theta'^H\mathbf B)\}-\mathrm{tr}(\bm\Phi'\mathbf A\bm\Phi'^H\mathbf B),
		\end{equation}
		where the equality is achieved if and only if $ \bm\Phi'=\bm\Theta' $.
	\end{lemma}
	\textit{Proof:}
	This can be easily verified   by the inequality $ \mathrm{tr}((\bm\Theta'-\bm\Phi')\mathbf A(\bm\Theta'-\bm\Phi')^H\mathbf B) \geq 0$ \cite{dongming2020}.
	$\hfill\blacksquare$

	Based on Lemma \ref{SCA_matrix}, we further introduce an auxiliary variable $ \bm\Phi $ to transform problem \eqref{P4Thetas} into:
	\begin{align}\label{P4ThetaPhi}
			\max_{\bm \Phi, \bm\Theta\in\mathcal M}\,\,& 2\Re\left\{\mathrm{tr}\left( \bm\Theta \mathbf M \right) \right\}+2\Re\left\{\mathrm{tr}\left(\bm\Phi\mathbf X\bm\Theta^H\left(\rho_1\mathbf I_N-\mathbf Y\right) \right) \right\} \notag  \\
			&\qquad\qquad\qquad-\mathrm{tr}\left(\mathbf \Phi\mathbf X\mathbf \Phi^H(\rho_1\mathbf I_N-\mathbf Y) \right).   
	\end{align}
	 In the following, we propose an AO algorithm to iteratively optimize $\bm \Phi$ and $\bm \Theta$ in \eqref{P4ThetaPhi} with closed form solutions. Specifically, with a given $\bm \Theta$, the optimal $\bm\Phi^\star$ is $ \bm\Phi^\star=\bm \Theta $ according to Lemma \ref{SCA_matrix}.

	For a given $\bm \Phi$, problem \eqref{P4ThetaPhi} becomes:
	\begin{align}\label{P4ThetaPhi-givenPhi}
			\max_{ \bm\Theta\in\mathcal M}\,\,& 2\Re\left\{\mathrm{tr}\left( \bm\Theta \mathbf M \right) \right\}+2\Re\left\{\mathrm{tr}\left(\bm\Phi\mathbf X\bm\Theta^H\left(\rho_1\mathbf I_N-\mathbf Y\right) \right) \right\}  \notag \\
			&\qquad\qquad\qquad-\mathrm{tr}\left(\mathbf \Phi\mathbf X\mathbf \Phi^H(\rho_1\mathbf I_N-\mathbf Y) \right).   
	\end{align}
	The optimal $\bm \Theta$ of problem \eqref{P4ThetaPhi-givenPhi} is obtained by the following definition and theorem.

	\begin{definition}\label{def:projection}
		For any square matrix $ \mathbf{Z} \in \mathbb{C}^{N\times N}$, let $\widehat{\mathbf{Z}}=\mathbf{Z}+\mathbf{Z}^T $ with a rank of $ R $. Its singular value decomposition (SVD) is $ \widehat{\mathbf{Z}}=\mathbf U \mathbf S \mathbf V^H $. Then, $ \mathbf U,\mathbf V $ can be partitioned into $ [\mathbf U_R, \mathbf U_{N-R} ] $ and $ [\mathbf V_R,\mathbf V_{N-R} ] $, respectively. Define a projection operator $ \bm\Pi_{\mathcal M}(\cdot):\mathbb{C}^{N\times N} \longrightarrow \mathcal M $ that projects any square matrix $ \mathbf{Z} $ to $\mathcal{M}$, as follows: 
		\begin{equation}\label{symuni}
			\bm\Pi_{\mathcal M}(\mathbf{Z})=\widehat{\mathbf U} \mathbf V^H,
		\end{equation}
		where $ \widehat{\mathbf U} \triangleq [\mathbf U_R, (\mathbf V_{N-R})^*]$.
	\end{definition}
	It has been proved in Proposition 1 of \cite{fang2023low} that the symmetric unitary projection defined in \eqref{symuni} is guaranteed to identify the point in $\mathcal M$ that is closest to $\mathbf{Z}$, i.e.,
	\begin{equation}
		\label{eq:proj_opt}
		\bm\Pi_{\mathcal M}(\mathbf{Z})=\arg \min_{\mathbf{Q}\in \mathcal M} \|\mathbf{Z}-\mathbf{Q}\|_F^2.
	\end{equation}



	\begin{theorem}\label{theo:optimal-projection}
		The optimal $ \bm \Theta $ for problem \eqref{P4ThetaPhi-givenPhi} is
		\begin{equation}\label{eq:opt-theta}
			\bm\Theta^\star = \bm\Pi_{\mathcal M}\left( \left(\rho_1\mathbf I_N-\mathbf Y\right) \bm \Phi \mathbf X + \mathbf M^H \right).
		\end{equation}		
	\end{theorem}
	\textit{Proof.}  
	Since $ \bm\Theta\bm\Theta^H=\mathbf I_N $, problem \eqref{P4ThetaPhi-givenPhi} is equivalent to:
	\begin{equation}\label{P4ThetaPhi-prove}
		\begin{split}
			\min_{ \bm\Theta\in\mathcal M} \left\| \bm \Theta - \left(\left(\rho_1\mathbf I_N-\mathbf Y\right) \bm \Phi \mathbf X + \mathbf M^H\right)\right\|^2_F,   
		\end{split}
	\end{equation}
	According to \eqref{symuni} and \eqref{eq:proj_opt}, it is easy to obtain that the optimal solution of \eqref{P4ThetaPhi-prove} is  \eqref{eq:opt-theta}, which completes the proof.
	$\hfill\blacksquare$
	
	
	Inspired by the block-wise closed-form solutions of \eqref{P4ThetaPhi} and the equivalence between \eqref{P4ThetaPhi} and \eqref{eq:sub-theta}, we propose the following projected successive linear approximation (PSLA) algorithm to solve  subproblem \eqref{P4Theta}:
\begin{itemize}
    \item  Step 1: Initialize $ \bm\Theta \in \mathcal{M}$;
    \item  Step 2: Compute $ \bm\Phi= \bm \Theta $;
    \item  Step 3: Compute $\bm\Theta=\bm\Pi_{\mathcal M}(\left(\rho_1\mathbf I_N-\mathbf Y\right)\bm\Phi\mathbf X + \mathbf M^H) $;
    \item  Step 4: Return to Step 2) until convergence.
\end{itemize}
	\subsection{Active Beamforming Optimization}\label{FP_W}
	When $ \bm\alpha,\bm\beta,\bm\Theta $ are fixed, the subproblem of \eqref{P4} with respect to $ \mathbf W $ is equivalent to\begin{equation}\label{eq:sub-w}
		\max_{\mathbf W\in\mathcal S }\,\, \sum_{k=1}^K2 \sqrt{1+\alpha_k} \delta_k \Re\{\beta_k^*\mathbf f_k^H\mathbf w_k \}
		-|\beta_k|^2 \sum_{j=1}^K|\mathbf f_k^H\mathbf w_j|^2.	
	\end{equation}
	By further defining $ \mathbf F\triangleq [\mathbf f_1,\cdots,\mathbf f_K]\in\mathbb C^{L\times K} $, problem \eqref{eq:sub-w} is equivalently transformed to:
	\begin{equation}\label{P4W}
		\max_{\mathbf W\in\mathcal S}\,\, 2\Re\{\mathrm{tr}\left(\mathbf \Sigma_1\mathbf F^H\mathbf W   \right)\}-\mathrm{tr}\left(\mathbf W\mathbf W^H\mathbf F\mathbf \Sigma_2\mathbf F^H\right).
	\end{equation}
	The objective function in \eqref{P4W} has a structure similar to that of \eqref{P4Theta}, it can be solved by the same PSLA framework as in Section \ref{FP_Theta}. Specifically, by introducing a shift parameter $\rho_2$, we transform \eqref{P4W} into:
	\begin{equation}\label{P4Wshift}
		\max_{\mathbf W\in\mathcal S}\,\, 2\Re\{\mathrm{tr}\left(\mathbf \Sigma_1\mathbf F^H\mathbf W   \right)\}+\mathrm{tr}\left(\mathbf W\mathbf W^H\left(\rho_2\mathbf I_L-\mathbf F\mathbf \Sigma_2\mathbf F^H\right)\right).
	\end{equation}
	By introducing an auxiliary matrix variable $ \mathbf{P} \in\mathbb C^{L\times K} $, problem \eqref{P4Wshift} is then transformed to:
	\begin{align}\label{P4Wshift-linear}
			\max_{\mathbf P, \mathbf W\in\mathcal S}\,\, &2\Re\{\mathrm{tr}\left(\mathbf \Sigma_1\mathbf F^H\mathbf W   \right)\} - \mathrm{tr}\left(\mathbf P\mathbf P^H\left(\rho_2\mathbf I_L-\mathbf F\mathbf \Sigma_2\mathbf F^H\right)\right) \notag\\
			&+2\Re\left\{\mathrm{tr}\left(\mathbf P\mathbf W^H\left(\rho_2\mathbf I_L-\mathbf F\mathbf \Sigma_2\mathbf F^H\right)\right)\right\}.
	\end{align}
	According to Lemma \ref{SCA_matrix}, with a given $\mathbf{W}$, the optimal $\mathbf{P}^{\star}$ is $ \mathbf{P}^\star = \mathbf{W} $. Similar to Theorem \ref{theo:optimal-projection}, with a given $\mathbf{P}$, the optimal $\mathbf{W}^{\star}$ is  
	\begin{equation}
		\mathbf{W}^\star=\bm\Pi_{\mathcal S}\left(\left(\rho_2\mathbf I_L - \mathbf F \bm \Sigma_2 \mathbf F^H\right) \mathbf P + \mathbf F \bm \Sigma_1 \right),
	\end{equation}
	where
	\begin{equation*}
		\begin{aligned}
			\bm \Pi_{\mathcal S}(\mathbf W)&\triangleq\arg\min_{\mathbf Q\in\mathcal S} \|\mathbf W-\mathbf Q\|_F^2=\sqrt{P_t}\mathbf W/\|\mathbf W\|_F.
		\end{aligned}
	\end{equation*}

	\subsection{Alternating Optimization for Problem \eqref{P1}}
	Based on the closed-form solution of $ \bm \alpha $ and $ \bm \beta $ obtained in Section \ref{FP}, and the iterative closed-form solutions of $\bm \Theta $ and $ \mathbf{W} $ in Section \ref{FP_Theta} and \ref{FP_W}, we propose an efficient AO algorithm, named FP-PSLA to address problem \eqref{P1}. The details of FP-PSLA are delineated in Algorithm \ref{alg:proposed}. Beginning with a non-zero feasible active beamforming matrix $ \mathbf W^{[0]} $ and a feasible symmetric unitary matrix $ \bm\Theta^{[0]} $, we iteratively compute the auxiliary variables $ \bm \alpha^{[n]} $ and $ \bm  \beta^{[n]} $ based on \eqref{eq:optalphabeta}, as well as the passive and active beamforming matrices $ \bm\Theta^{[n]},\mathbf W^{[n]} $ based on the proposed PSLA algorithm until the WSR converges.

 	\setlength{\textfloatsep}{7pt}	
	\begin{algorithm}[t!]
		\caption{Proposed FP-PSLA algorithm to solve problem (\ref{P1})}
		\label{alg:proposed}
		\textbf{Input:} The channel matrices $ \mathbf G,\mathbf H $ and $ \mathbf E $, maximum transmit power budget $ P_t $, convergence tolerance $ \epsilon $\;
		\textbf{Initilization:} $ n\leftarrow 0 $, $\mathbf W^{[0]}, \bm\Theta^{[0]}$\;
		\Repeat{$|\mathrm{WSR}^{[n]}-\mathrm{WSR}^{[n-1]}|\leq \epsilon$}{ 
			$ n\leftarrow n+1 $\;
			Update $ \alpha_{k}^{[n]} $ and $ \beta_{k}^{[n]} $ by \eqref{eq:optalpha} and \eqref{eq:optbeta}, $ \forall k\in\mathcal{K} $\;
			\textbf{Initilization:} $\mathbf W^{[m]}\leftarrow\mathbf W^{[n-1]}$ and $m=0$;\\
			\Repeat{convergence}{ 
				$ \mathbf P^{[m]}= \mathbf{W}^{[m]} $ \;
				$ \mathbf W^{[m+1]}=\bm\Pi_{\mathcal S}\left(\left(\rho_2\mathbf I_L - \mathbf F \bm \Sigma_2 \mathbf F^H\right) \mathbf P^{[m]} + \mathbf F \bm \Sigma_1\right) $\;
				$ m\leftarrow m+1 $\;
			}
			\textbf{Initilization:} $\bm \Theta^{[t]}\leftarrow \bm \Theta^{[n-1]}$ and $t=0$;\\
			\Repeat{convergence}{ 
				$ \bm\Phi^{[t]}=\bm \Theta^{[t]} $ \;
				$ \bm\Theta^{[t+1]}=\bm\Pi_{\mathcal M}(\left(\rho_1\mathbf I_N-\mathbf Y\right)\bm\Phi^{[t]}\mathbf X + \mathbf M^H) $\;
				$ t\leftarrow t+1 $\;
			}
		}
	\end{algorithm}
	%

{\subsection{Convergence  Analysis}}
{
The proposed FP-PSLA algorithm involves two loops: the outer loop and the inner loop. The convergence analysis of the outer loop follows \cite{shen2018fractional,fang2024ratesplittingmultipleaccess}. In this subsection, we delve into the convergence of the inner loop. Let the objective function of problem \eqref{P4ThetaPhi-givenPhi} be denoted as $g(\bm \Theta, \bm\Phi)$ and the objective function of the original problem \eqref{P4Theta} be denoted as $f(\bm \Theta)$. According to Lemma \ref{SCA_matrix}, when $ \bm\Phi^{[t]} = \bm \Theta^{[t]} $ in the inner loop iteration $[t]$, we have $g(\bm \Theta^{[t]},\bm\Phi^{[t]}) = f(\bm \Theta^{[t]})$  and $g(\bm \Theta^{[t+1]},\bm\Phi^{[t]}) \leq f(\bm \Theta^{[t+1]})$.  Since $\bm \Theta^{[t]}$ is a feasible solution   and $\bm \Theta^{[t+1]}$ is the optimal solution of \eqref{P4ThetaPhi-givenPhi} in each inner-loop iteration $[t+1]$, according to Theorem \ref{theo:optimal-projection}, we have $g(\bm \Theta^{[t+1]},\bm\Phi^{[t]}) \geq g(\bm \Theta^{[t]},\bm\Phi^{[t]})$. Therefore, $f(\bm \Theta^{[t+1]}) \geq g(\bm \Theta^{[t+1]},\bm\Phi^{[t]}) \geq g(\bm \Theta^{[t]},\bm\Phi^{[t]}) = f(\bm \Theta^{[t]})$. The objective function $f(\bm \Theta)$ is monotonically increasing as the iteration increases.  
Given that the solution to problem \eqref{P4Theta} is constrained within a finite set $\mathcal M$, the convergence of the passive beamforming optimization in the inner loop of the proposed FP-PSLA algorithm is guaranteed. This analysis  also applies to the active beamforming optimization.   Therefore, the proposed FP-PSLA algorithm is guaranteed to converge within a given tolerance $\epsilon$.


	\section{Numerical Results}

        \begin{figure*}[htbp]
            \centering
            \subfloat[\label{fig:convergence}Convergence of the proposed FP-PSLA algorithm.]
            {\includegraphics[width=0.3\textwidth]{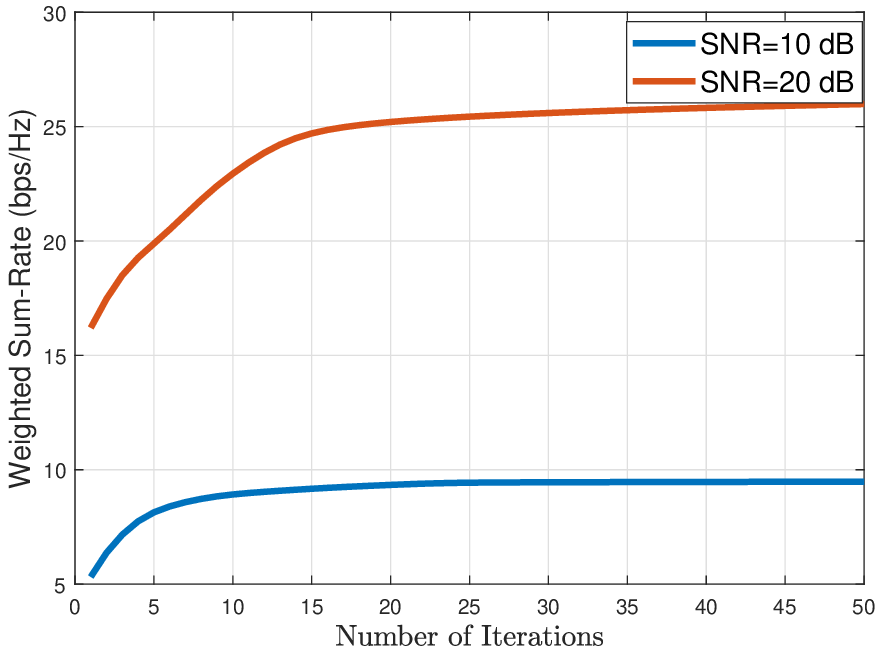}}
            \hfill
            \subfloat[\label{fig:WSR-N}WSR versus the number of BD-RIS elements $ N $.]
            {\includegraphics[width=0.3\textwidth]{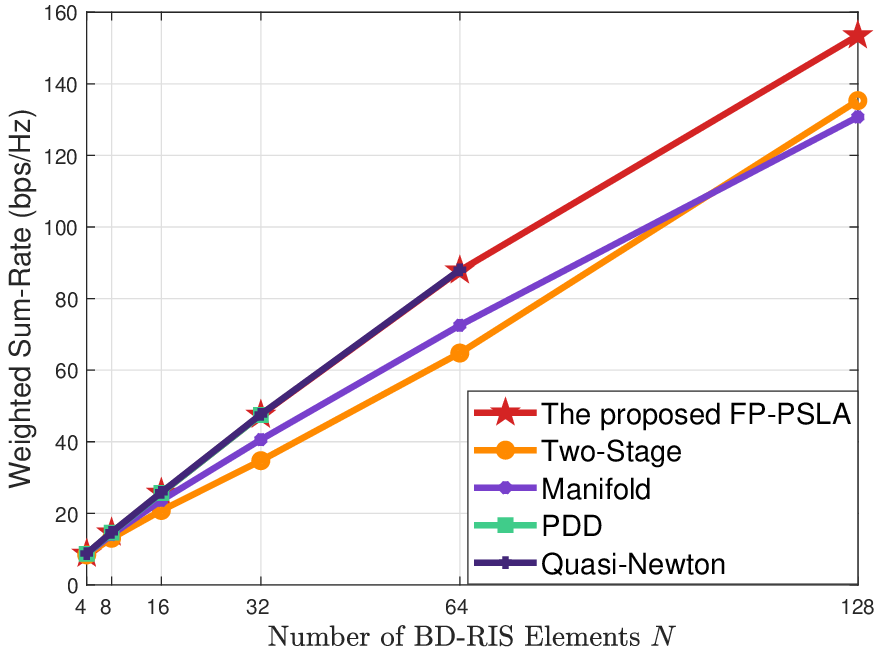}}
            \hfill
            \subfloat[\label{fig:T-N}Average CPU time versus the number of BD-RIS elements $ N $.]
            {\includegraphics[width=0.3\textwidth]{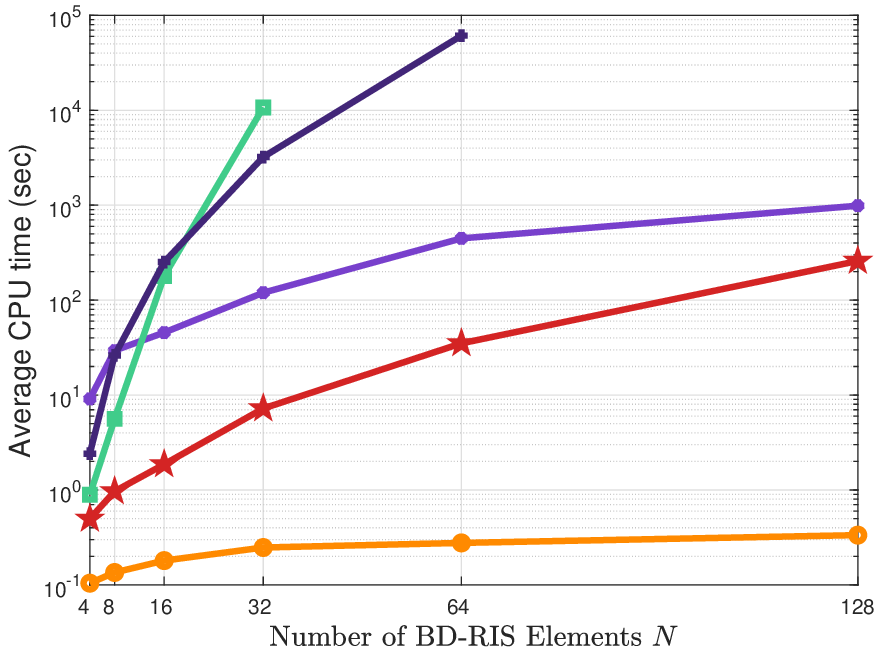}}

            \caption{Analysis of the proposed algorithm.}
        \end{figure*}

	In this section, the simulation results of the proposed algorithm are demonstrated. We assume that the BS and RIS are positioned at coordinates (0,0) and (150,50) meters respectively. Additionally, we consider a scenario with $ K=32 $ users randomly distributed within a circular area centered at (150,0) meters, with a diameter of 20 meters and the number of transmit antennas is $ L = 32 $. The path loss is  $ P(d)=L_0 d^{-\alpha} $, where $ L_0 = -30 $ dB denotes the reference path loss at $ d=1 $ m, $ d $ represents the link distance, and $ \alpha $ is the path loss exponent. Following \cite{fang2022fully}, the path loss exponents from the BS to RIS and RIS to users are 2 and 2.2, respectively. 
	The small-scale fading model follows Rayleigh fading. The convergence tolerance is $ \epsilon = 10^{-3} $.
	All simulation results are averaged over 100 random channel realizations.

	\par We compare the proposed FP-PSLA with the following four baselines to solve problem \eqref{P1}: the quasi-Newton Algorithm in \cite{shen2021modeling,fang2022fully}, the {Manifold Algorithm} in \cite{hongyu2023,hongyu2024,hongyu2023jasc}, the {PDD Algorithm} in \cite{yuyuan2023}, and the {Two-Stage Algorithm} in \cite{fang2023low}. 
	For all algorithms, the active beamforming matrix is initialized using the maximum ratio transmission (MRT) as $ \mathbf{W}^{[0]}=\sqrt{0.4P_t} \mathbf{F}/\|\mathbf{F}\|_F $, while the passive beamforming matrix initialization follows \cite{fang2023low}, 
	i.e., $\bm \Theta^{[0]}=\bm\Pi_{\mathcal M}( \mathbf H\mathbf{I}_{K \times N}\mathbf E^H) $.	
	Table \ref{tab:complexity} provides the  computational complexity comparison among all algorithms, where $ I_{\bm\Theta}$, $I_{\mathbf W} $ represent the number of iterations for each inner loop and $ I_t $ refers to the number of iterations for the outer loop. Except the heuristic Two-Stage algorithms, the proposed FP-PSLA algorithm has the lowest computational complexity compared to other optimization algorithms.

	\begin{table}[t!]
		\begin{center}
			\caption{Computational Complexity Comparison for Different Algorithms to Solve Problem \eqref{P1}.}
			\label{tab:complexity}
			\begin{tabular}{ccc}
				\toprule
				\textbf{Algorithm}& 	\textbf{Reference} & \textbf{Computational Complexity}  \\
				\hline\hline
				Quasi-Newton & \cite{fang2022fully}&$ \mathcal O\left(I_t\left(I_{\bm \Theta}N^4+I_{\mathbf{W}}(KL)^{3.5}\right)\right)$  \\
				\hline
				Manifold & \cite{hongyu2023}&$\mathcal O(I_t(I_{\bm \Theta}N^3+I_{\mathbf{W}}KL^3)) $  \\
				\hline
				PDD & \cite{yuyuan2023} &$\mathcal O(I_t(I_{\bm \Theta}N^6+I_{\mathbf{W}}KL^3)) $\\
				\hline
				Two-Stage & \cite{fang2023low}&$\mathcal O(N^3+I_{t}I_{\mathbf{W}}KL^3) $\\
				\hline
				FP-PSLA &Proposed& $ \mathcal O( I_{t}(I_{\bm\Theta}N^3+I_{\mathbf W}L^2K ) ) $\\
				\bottomrule
			\end{tabular}
		\end{center}
	\end{table}

	Fig. \ref{fig:convergence} illustrates the convergence behavior of the proposed FP-PSLA algorithm under various transmit signal-to-noise ratio (SNR) when $ N=16 $. It is evident that the proposed algorithm converges within a few outer-loop iterations.	
	

	Fig. \ref{fig:WSR-N} and \ref{fig:T-N} respectively illustrate the WSR performance and the average CPU time in relation to the number of BD-RIS elements for different algorithms. The transmit SNR is $ 20 $ dB. In Fig. \ref{fig:WSR-N}, it is notable that the proposed FP-PSLA algorithm achieves comparable performance to both the quasi-Newton and PDD algorithms, while outperforming the Manifold and Two-Stage algorithms. Specifically, the proposed FP-PSLA algorithm outperforms the Two-Stage algorithm  by a relative WSR gain of 20.82\% on average and the Manifold algorithm by a relative WSR gain of 11.48\% on average. Most importantly, according to  Fig. \ref{fig:T-N}, among the algorithms that achieve the near optimal performance, the proposed FP-PSLA algorithm consumes the least average CPU time. Specifically, the proposed FP-PSLA algorithm demonstrates an average CPU time reduction of 91.19\% compared to the Manifold algorithm, 
	and 94.94\% compared to the quasi-Newton algorithm. 
	Hence, the proposed FP-PSLA algorithm substantially decreases the CPU time while ensuring a near optimal solution.
	%
	%
	%
	\section{Conclusion}
	In this paper, for a downlink BD-RIS-assisted multi-user multiple-input multiple-output transmission network, we propose a novel efficient optimization algorithm, named FP-PSLA, to jointly optimize the active and passive beamforming matrices in order to maximize the WSR. The primary novelty of the proposed algorithm lies in a unified optimization framework that relies on iterative closed-form solutions to solve the subproblems of active and passive beamforming design. Numerical results show that the proposed FP-PSLA algorithm significantly reduces the CPU time while maintaining the same performance as other baseline schemes that guarantee near optimal solutions. Therefore, we conclude that the proposed algorithm holds great potential for the application of BD-RIS in practice. Future work will consider joint beamforming design, accounting for hardware impairments and more realistic RIS models \cite{hardware}.
	


	\bibliographystyle{IEEEtran}  
	\bibliography{reference}

\end{document}